\documentclass[%
reprint,prl,
superscriptaddress,
amsmath,amssymb,
aps,
]{revtex4-1}

\usepackage{xparse}
\usepackage{braket}

\usepackage{mathtools}
\usepackage{appendix}
\usepackage{graphicx}
\usepackage{dcolumn}
\usepackage{bm}
\usepackage{xcolor}
\usepackage[normalem]{ulem}
\usepackage{natbib}

\usepackage{float}
\usepackage{braket}
\usepackage{bbold}
\usepackage{mathtools}

\DeclareMathOperator{\Tr}{Tr}
\DeclareMathOperator{\tr}{tr}

\DeclareMathOperator{\diag}{diag}

\newcommand{\msf}[1]{\mathsf{#1}}

\newcommand{\w}{\omega}

\newcommand{\s}{e^{-\sigma^2}}
\newcommand{\st}{e^{-t\sigma^2}}

\begin{document}
	
	\title{
		Slow Relaxation in a Glassy Quantum Circuit
	}
    \author{Richard D. Barney}
    \affiliation{Joint Quantum Institute, Department of Physics, University of Maryland, College Park, Maryland, USA}
      \author{Yunxiang Liao}
    \affiliation{Department of Physics, KTH Royal Institute of Technology, Stockholm, Sweden}
    \author{Victor Galitski}
    \affiliation{Joint Quantum Institute, Department of Physics, University of Maryland, College Park, Maryland, USA}
    
    \begin{abstract}
    Quantum circuits have become a powerful tool in the study of many-body quantum physics, providing insights into both fast-thermalizing chaotic and non-thermalizing integrable many-body dynamics. In this work, we explore a distinct intermediate class---glassy quantum systems---where thermalization occurs, but over very long timescales. We introduce and analyze a Floquet random quantum circuit that can be tuned between glassy and fully ergodic behavior through a single adjustable parameter. This circuit can be understood as the unitary analog of the block Rosenzweig-Porter model, which is defined by a Hamiltonian. Using an effective field theory for random quantum circuits, we analyze the correlations between quasienergy eigenstates and thereby determine the time evolution of the disorder-averaged density matrix. In the intermediate regime the circuit displays a two-step thermalization process: an initial relaxation within weakly coupled sectors followed by a later, global thermalization. We also show that the ramp of the spectral form factor is enhanced by a factor of the number of sectors in the glassy regime, and at early times in the intermediate regime.
    These results indicate that quantum circuits provide an ideal platform for the exploration of nontrivial thermalization dynamics in many-body quantum systems, offering deeper insights into quantum thermalization.
    \end{abstract}   


	\maketitle

 
	Understanding which type of isolated quantum interacting systems, when starting from an out-of-equilibrium initial state,  eventually relax to thermal equilibrium is a foundational question for the validity and applicability of quantum statistical mechanics. This question is partially addressed by the eigenstate thermalization hypothesis (ETH)~\cite{Deutsch1991,Srednicki1994,DAlessio2016,Deutsch_2018}, which suggests that quantum chaotic systems, characterized by particular energy eigenstate statistics, can achieve thermal equilibrium under their own dynamics. Compared with earlier attempts which try to explain thermalization in isolated quantum systems, ETH can distinguish systems that can thermalize from those that cannot, based on different statistical properties of their energy eigenstates.
	
	ETH may break down in certain types of quantum systems, including integrable systems with an extensive number of conserved quantities~\cite{Rigol2007,Rigol2008,Rigol2009,Polkovnikov2011,Eisert2015,Gogolin2016} and many-body localized (MBL) systems where strong disorder prevents thermalization even in  the presence of interactions~\cite{Altshuler1997,Basko2006,Imbrie2016,Huse2013,Rahul2015,Balasubramanian2020}.
	While there is a substantial number of numerical~\cite{MBL-Poisson,Pal2010,Iyer2013,Luitz2015,Mondaini2015,Sierant2018,Orell2019,Mace2019,Hopjan2020} and experimental~\cite{Yao2014,Schreiber2015,Choi2016,Smith2016,Zhang2017,Bordia2017,Bordia2017_periodically,Luschen2017,Luschen2017_observation,Wei2018} studies supporting MBL in small size systems, the existence of MBL in the thermodynamic limit remains a topic of active debate~\cite{deRoeck2016,Panda2019,Suntajs2020,Sierant2020,Abanin2021,Sierant2022}.
	Additionally, there exists a special type of non-ergodic systems where thermalization can occur, but at an extremely slow rate: glasses~\cite{Sherrington1975Solvable,Edwards1975,Binder1986,Mezard1987,Goldschmidt1990,Wenhao1991,Fischer1991,Rieger1997,Cugliandolo2001,Castellani2005,Stein2013,Winer2022}. The slow relaxation dynamics in these systems remain relatively less explored, particularly concerning the the underlying microscopic mechanism.

	Wigner-Dyson energy level statistics~\cite{Wigner_1951,Dyson1962_statistical} has been widely used as a defining property of quantum chaotic systems \cite{Sieber_2001,Muller_2005,Chalker-1,Prosen-SFF1,SSS,PRR,PRB}, based on the Bohigas-Giannoni-Schmit conjecture~\cite{Bohigas1984}. By contrast, integrable systems and MBL systems instead exhibit Poisson statistics \cite{Berry-Tabor,MBL-Poisson}.
	Recently, it has been demonstrated that a particular quantum spin glass model, the quantum p-spherical model, exhibits energy level statistics distinct from those of quantum chaotic systems and of integrable systems~\cite{Winer2022}. Its statistical properties are analogous to that of a random matrix model which falls outside the Wigner-Dyson class, known as the block Rosenzweig-Porter model~\cite{Barney2023}. 
	In particular, the spectral form factors (SFFs) \cite{BerrySFF,Zoller}---a diagnostic of energy level statistics---of both the quantum p-spherical model and the block Rosenzweig-Porter model exhibit a linear-in-time ramp, analogous to Wigner-Dyson statistics, but with a coefficient that is enhanced by the number of weakly coupled sectors (metastable configurations).
	
	Quantum circuits have become a widely used platform for the study of many-body quantum physics and can be implemented in digital quantum simulators in various experimental platforms~\cite{Lewenstein2007,Gross2017,Schafer2020,Lanyon2011,Blatt2012,Wendin2017,Kjaergaard2020,Altman2021}. In particular, some of their fine-tuned properties, such as dual unitarity~\cite{Bertini2018,Bertini2019,Bertini2019_entanglement,Gopalakrishnan2019} or the randomness of the quantum gates~\cite{Oliveira2007,Znidaric2008,Hamma2012,Nahum2017,Nahum2018,vonKeyserlingk2018,Zhou2019,Fisher2023,Chalker-1,Chalker-2}, allow for exact theoretical investigation of the thermalization dynamics as well as other chaotic behaviors which also occur in more realistic quantum chaotic systems. This makes quantum circuits an ideal candidate for the exploration of ergodicity breaking and realization of glassy phases characterized by extremely slow relaxation processes,
	leading to a deeper understanding of the fundamental question of quantum thermalization.
	
	In this paper, we propose a Floquet random quantum circuit, a random circuit with discrete time-translation symmetry, which can exhibit slow glassy thermalization dynamics.
	The Floquet operator of this quantum circuit can be considered as a unitary version of the block Rosenzweig-Porter model.
	The quasienergy eigenstates and eigenvalues of this glassy quantum circuit possess statistical properties distinct from those of quantum chaotic systems. Its relaxation towards thermal equilibrium involves two steps:  first thermalization occurs within different sectors that are weakly coupled to each other; then this is  followed by global thermalization across the entire Hilbert space.
	Moreover, by tuning a parameter in the model which represents the strength of the coupling between different sectors, the rate of global thermalization can be made to be extremely slow.
	
	\begin{figure}[t]
		\centering
		\includegraphics[width=0.95\linewidth]{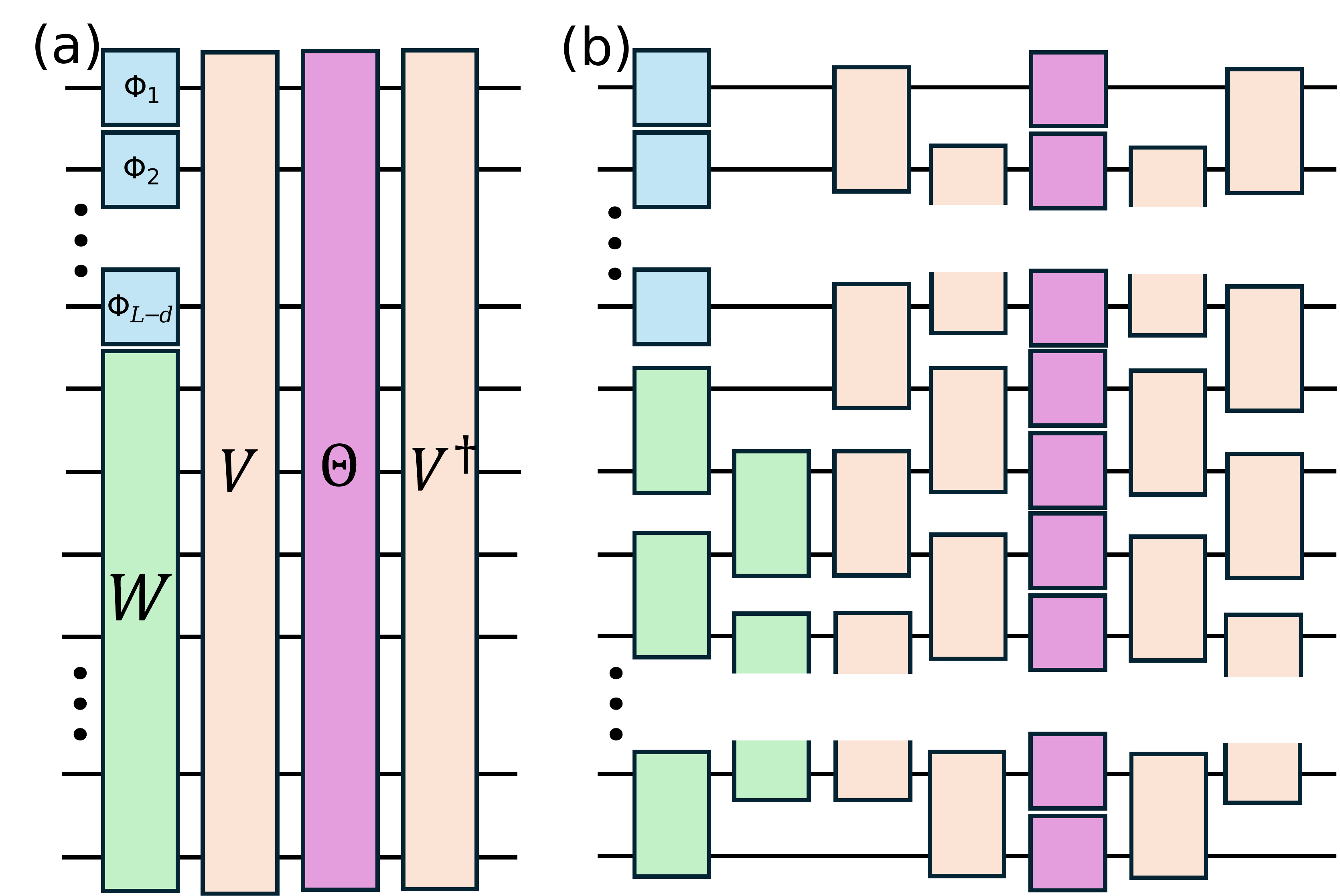}
		\caption{ 
			(a) The time periodic glassy random quantum circuit's Floquet operator.
			The first layer involves independent random phase gates $\Phi_i$ acting individually on the first $L-D$ qudits,  together with a CUE random matrix $W$ acting on the remaining $D$ qudits. The subsequent layers include an independent CUE matrix $V$, a global quantum phase gate $\Theta$, and $V^{\dagger}$, all acting on the entire chain of qudits.
			The random phase gates $\Phi$ and $\Theta$ are diagonal matrices with dimensions $q$  and $q^L$, respectively. 
			The phases in $\Phi$ are independently and uniformly distributed over $(-\pi,\pi]$, while those in  $\Theta$ are independently Gaussian distributed with zero mean and variance $\sigma^2$. 
			By tuning $\sigma$, the process of thermalization  can be significantly slowed down, and it is completely prevented in the $\sigma \rightarrow 0$ limit, in which case only the first layer is effective.
	(b) A schematic representation of the Floquet operator implemented with local gates. In practice, the global operators $W$, $V$ and $\Theta$ can be implemented using layers of local quantum gates that act on individual or neighboring qudits, as long as the effective time evolution operators exhibit the same statistical properties.
		}
		\label{fig:f1}
	\end{figure}

    
	The  Floquet quantum circuit under consideration consists of a one-dimensional  chain of $L$ qudits, each with $q$ internal levels. For our analytic calculation, we focus on the large $q$ limit. The qudits evolve under discrete  and periodic applications of unitary quantum gates depicted in Fig.~\ref{fig:f1}(a). In particular, the Floquet operator $U$  takes the form of
	\begin{align}\label{eq:U}
		U= V^\dagger \Theta V \left(\bigotimes_{i=1}^{L-D}  \Phi_i \otimes W\right) .
	\end{align}
	Here $\Phi_i\equiv \diag{(e^{i\phi_1^{(i)}},...,e^{i\phi_q^{(i)}})}$ represents a random phase gate acting on a single qudit at site $1 \leq i\leq L-D$, with the diagonal matrix element $\phi_j^{(i)}$ being independently and uniformly distributed within the range $(-\pi,\pi]$. 
	$\Theta=\diag{(e^{i\theta_1},...,e^{i\theta_{q^L}})}$ is another random phase gate that acts on all qudits, where each random phase $\theta_j$ follows an independent Gaussian distribution with zero mean and variance $\sigma^2$.
	$W$ ($V$) is a Haar-distributed $M \times M$ ($N
	\times N$) random unitary matrix acting on  sites $ [L-D+1, L]$ (all sites), where $M \equiv q^D$ ($N \equiv q^L$) is the Hilbert space of $D$ ($L$) qudits.
	In practice, the random unitary gates $W$ and $V$ can be constructed from layers of two-qudit unitary gates, such that the statistical properties of the resulting effective time evolution operators match those of the circular unitary ensemble (CUE) of the same dimensions~\cite{Mehta}. Similarly, the global random phase gate $\Theta$ can also be generated from a series of single-qudit random phase gates (Fig.\ref{fig:f1}(b)).

	
	The time evolution operator for the first layer of quantum gates $\left(\bigotimes_{i=1}^{L-D}  \Phi_i \otimes W\right) =\bigoplus_{i=1}^{P} A^{(i)} $ has a block diagonal structure with each diagonal block being an $M \times M$ matrix of the form
	$A^{(i)}=e^{ i\phi_{i}'}W$. Here $P$ is defined as $P \equiv N/M$. All phases $\phi_{i}'$ are given by sums of $L-D$ random phases from the set $\left\lbrace\phi_{j}^{(l)}|j=1,2,...,q;l=1,2,...,L-D\right\rbrace$. 
	Using the fact that  $\phi_{j}^{(l)}$ are  independently and uniformly distributed, one can show that the correlation $\left\langle A^{(l)}_{ij}(A^{(l')})^{\dagger}_{j'i'} \right\rangle 
	=\delta_{ll'}\delta_{ii'}\delta_{jj'}/M$, identical to the correlation of statistically independent CUE matrices of dimension $M$. Here we have used the angular bracket to indicate ensemble averaging. 
	In the following, we approximate $A^{(l)}$, $l=1,2,...,P$, by statistically independent CUE matrices of dimension $M$ to simplify the calculation, believing this approximation effectively captures the essential relaxation behavior of the glassy quantum circuit under consideration. 
	
	
	The remaining layers of quantum gates  $V\Theta V^\dagger$ couple the diagonal blocks $A^{(i)}$, with the coupling strength tuned by the variance $\sigma^2$ of the random phase gate $\Theta$.
	When $\sigma = 0$, $\Theta$ becomes an identity matrix, making the Floquet operator $U=\bigoplus_{i=1}^{P} A^{(i)} $ block diagonal. In the opposite limit $\sigma \gg 1$, the coupling between $A^{(i)}$ becomes so strong that we expect the random circuit to be quantum chaotic with Wigner-Dyson statistics. In the intermediate regime where $\sigma \sim O(1)$, the Floquet operator $U$ possesses a structure analogous to that of the block Rosenzweig-Porter model in the intermediate regime, which serves as a minimal model for quantum spin glasses. 
	Therefore, similar glassy behaviors in its relaxation dynamics are expected in this intermediate regime.

	
	The correlation of the Floquet operator $U$ is given by,
	\begin{align}\label{eq:U2}
		\begin{aligned}
			\left\langle U_{ij} U^{\dagger}_{j'i'}\right\rangle 
			= 
			\delta_{ii'}\delta_{jj'}
			\left(
			\delta_{B(i),B(j)}
			\frac{1+Ne^{-\sigma^2}}{M(N+1)} 
			+
			\frac{1-e^{-\sigma^2}}{N+1}
			\right).
		\end{aligned}
	\end{align}
	Here $B(i)\equiv \lceil i/M \rceil $ denotes the block index to which the integer $i$ belongs when the $N\times N$ matrix is divided into $P\times P$ blocks, with each block being an $M\times M$ matrix. 
    When $\sigma = 0$, this correlation function reduces to that of a block
    CUE,  i.e., an $N\times N$ block diagonal matrix with statistically independent CUE matrices of dimension $M$  as its diagonal blocks.
	In the large $ \sigma $ limit the first term becomes much smaller than the second term, which reduces the correlation function to that of a CUE matrix. However, due to the presence of the Kronecker delta, the first term cannot be simply ignored when investigating the relaxation behavior. 

	To investigate the statistical properties of the quasienergy eigenstates and eigenvalues  of the Floquet random quantum circuit as well as its relaxation behaviors, we compute the following correlation function of the quasienergy eigenstates
	\begin{align}\label{eq:Comega}
		C_{nn'm'm}(\omega)
		=&
		\left\langle 
		\sum_{\mu,\nu}
		\psi_{n}^{\mu} (\psi_{n'}^{\mu})^*
		\psi_{m'}^{\nu} (\psi_{m}^{\nu})^*
		\delta_{2\pi} \left( \omega- E_\mu+E_{\nu}\right) 
		\right\rangle.
	\end{align}
	Here $\psi_{n}^{\mu}$ denotes the $n^\text{th}$ component of the quasienergy eigenstate with quasienergy $E_{\mu}$, and $ \delta_{2\pi}$ represents the $2\pi$-periodic Dirac delta function.
	The Fourier transform of $  C_{nn'm'm}(\omega)$ is  the correlation function of the matrix elements of the time evolution operator $U(t)=U^t$,
	\begin{align}\label{eq:Ct-0}
		\begin{aligned}
			&	C_{nn'm'm}(t)
			=
			\left\langle 
			U_{nn'}(t)
			U^{\dagger}_{m'm}(t)
			\right\rangle.
		\end{aligned}
	\end{align}
	From this eigenstate correlation function one can obtain the time evolution of the density matrix $\rho(t)$ starting from an arbitrary initial state $\rho(0)$:
	\begin{align}\label{eq:rhot-0}
		\left\langle \rho_{nm}(t) \right\rangle = \sum_{n'm'} C_{nn'm'm}(t) \rho_{n'm'}(0),
	\end{align}
	as well as the SFF:
	\begin{align}\label{eq:K}
		K(t) \equiv \left\langle 
		|\Tr U(t)|^2
		\right\rangle 
		=\sum_{nm} C_{nnmm}(t) .
	\end{align}

	To compute the eigenstate correlation function  for the Floquet quantum circuit under consideration, we employ a sigma model \cite{Efetov,Haake,Altland-rev} approach developed in Refs.~\cite{Liao2022,arxiv}. Since the detailed calculation is analogous to that in Ref.~\cite{arxiv} for a family of chaotic quantum circuits, we leave the full derivation to the Supplementary Material~\cite{Sup}.
	By focusing on the quadratic fluctuations around the standard saddle point in the sigma model, we obtain the smoothed eigenstate correlation function  $C_{nn'm'm}(\omega)$, where the $2\pi-$periodic Dirac delta function in Eq.~\ref{eq:Comega} is replaced by a sharp Lorenzian $L_{\eta}(\w) \equiv \sum_{n=-\infty}^{\infty} e^{i\w n-|n|\eta}/2\pi $ of width $M^{-1}\ll\eta\ll 1$~\cite{Altland-rev}.
	This smoothed correlation function
	measures the overlap of pairs of the quasienergy eigenstates $\mu, \nu$ whose energy separation $E_{\mu}-E_{\nu}$ is close to $\w$, weighted by the Lorenzian $L_{\eta}(\w-E_{\mu}+E_{\nu})$.
	Note that both $\eta$ and $1/q$ serve as the small parameters in our perturbative calculation \cite{wegner,wegner1980,Altland-rev}. To calculate the non-smoothed correlation function, which reveals the structure of nearby quasienergy eigenstates, or to study the finite $q$ case, it becames important to consider the fluctuations beyond the quadratic order.  We leave this investigation for future work.

	The smoothed quasienergy eigenstate correlation function takes the form:
	\begin{align}\label{eq:Ct}
		\begin{aligned}
			&{C}_{nn'm'm} (t)
			=
			\delta_{nn'}\delta_{mm'} \left(\delta_{t,0}+c_1(t) \right)
			+
			\delta_{nm}\delta_{n'm'}
			c_2(t)
			\\
			&
			+
			\delta_{nn'}\delta_{mm'}
			\delta_{B(n),B(m)}
			c_3(t)
			+
			\delta_{nm}\delta_{n'm'}
			\delta_{B(n),B(n')}
			c_4(t),
		\end{aligned} 
	\end{align}
	where, to the leading order,
	\begin{align}\label{eq:ct}
		\begin{aligned}
			&
			c_1(t)
			=
			\frac{1}{N^2}
			\left[
			t(1+e^{-t\sigma^2})
			-
			\frac{1+e^{-\sigma^2}  }{1-e^{-\sigma^2} }
			(
			1
			-
			e^{-t\sigma^2}
			)
			\right] ,
			\\
			&
			c_2(t)
			=
			\frac{1}{N}
			\left(
			1
			-
			e^{-t\sigma^2}
			\right),
			\\
			&
			c_3(t)
			=
			\frac{1}{N^2}
			\left[
			\frac{2P\s}{1-\s} \left(1-\st \right)
			+
			(P^2-2P)
			t
			\st
			\right.
			\\
			&
			\qquad\qquad-P^2
			(\st-\delta_{t,0})
			\Bigg],
			\\
			&	
			c_4(t)
			=
			\frac{1}{M} \left(\st-\delta_{t,0}\right)
			.
		\end{aligned}
	\end{align}
	See Eq.~S17 in the Supplementary Material~\cite{Sup} for the corresponding expression  in frequency space.
	This result is derived for the regime $\sigma \ll \sqrt{\ln N}$, but it can be extended to the regime of large $\sigma \gtrsim \sqrt{\ln N}$, where the eigenstate correlation function becomes that of the CUE. 
	Additionally, we emphasize that this expression, obtained from the Fourier transform of the smoothed eigenstate correlation function in frequency space,
 applies only at times much shorter than the block Heisenberg time, which is of the order $O(M)$.
	
	\begin{figure}[t]
		\centering
		\includegraphics[width=0.95\linewidth]{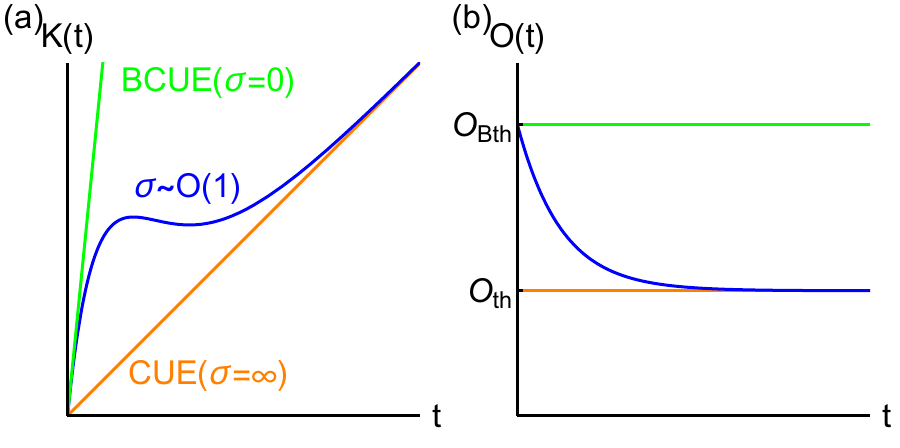}
		\caption{ 
			The glassy quantum circuit's (a) smoothed SFF $K(t)$  (Eq.~\ref{eq:Kt}) and (b) the time dependence of the expectation value for a physical observable $O(t)$ in the intermediate time regime (Eq.~\ref{eq:Ot}). Panel (a)  shows the SFF in the intermediate $\sigma$ regime (blue curve), which crosses over from the enhanced ramp for block CUE associated with $\sigma  \rightarrow 0$ (green curve) at early times to the standard ramp of CUE corresponding to $\sigma \rightarrow \infty $ (orange curve)  at later times. Panel (b) illustrates the relaxation of the expectation value of physical observable $O(t)$  from the block thermalized value $O_{\msf{Bth}}$ to the  fully thermalized value $O_{\msf{th}}$ using Eq.~\ref{eq:rhot}. The relaxation rate is also given by $\sigma$.
			Note here we show only the regime where the terms involving $\rho(0)$ and 
			$\rho_{\msf{B}}(0)$, which are important for the early relaxation from   $O(0)$ to $O_{\msf{Bth}}$, are negligible.
		}
		\label{fig:f2}
	\end{figure}


	The smoothed SFF can be derived from Eqs. \ref{eq:K} and \ref{eq:Ct}:
	\begin{align}\label{eq:Kt}
		\begin{aligned}
			&K(t)
			=
			e^{-\sigma^2t} Pt +  (1-e^{-\sigma^2t}) t
			+ N^2\delta_{t,0}.
		\end{aligned}
	\end{align} 
	As expected, this result reduces to that of the CUE when $\sigma \rightarrow \infty$,  and to that of the block CUE when $\sigma \rightarrow 0$.
	In both cases, the SFF exhibits a linear-in-$t$ ramp, but the ramp in the latter case is enhanced by a factor of the number of blocks $P$. 
	This enhancement is straightforward to understand since the SFF for the block CUE is simply given by the summation of the SFFs for each individual CUE block.
	Note that the plateau in the non-smoothed SFF occurs at large time and therefore cannot be recovered from the current perturbative calculation.
	In the intermediate $\sigma$ regime, the SFF of the current model interpolates between the limits of $\sigma\rightarrow 0$ and $\sigma\rightarrow \infty$, as shown in Fig.~\ref{fig:f2}(a).
	Specifically, it crosses over from the enhanced ramp of the block CUE at small times to the standard ramp of the CUE at larger times, with the rate  determined by $\sigma$.
	This SFF structure resembles that of the block Rosenzweig-Porter model \cite{Barney2023}.
	Notably, a similar enhanced ramp is observed in the SFF of the quantum p-spherical model~\cite{Winer2022}, with the enhanced factor $P$ being the number of metastable states.

\begin{figure}
      \centering
      \includegraphics[width=0.95\linewidth]{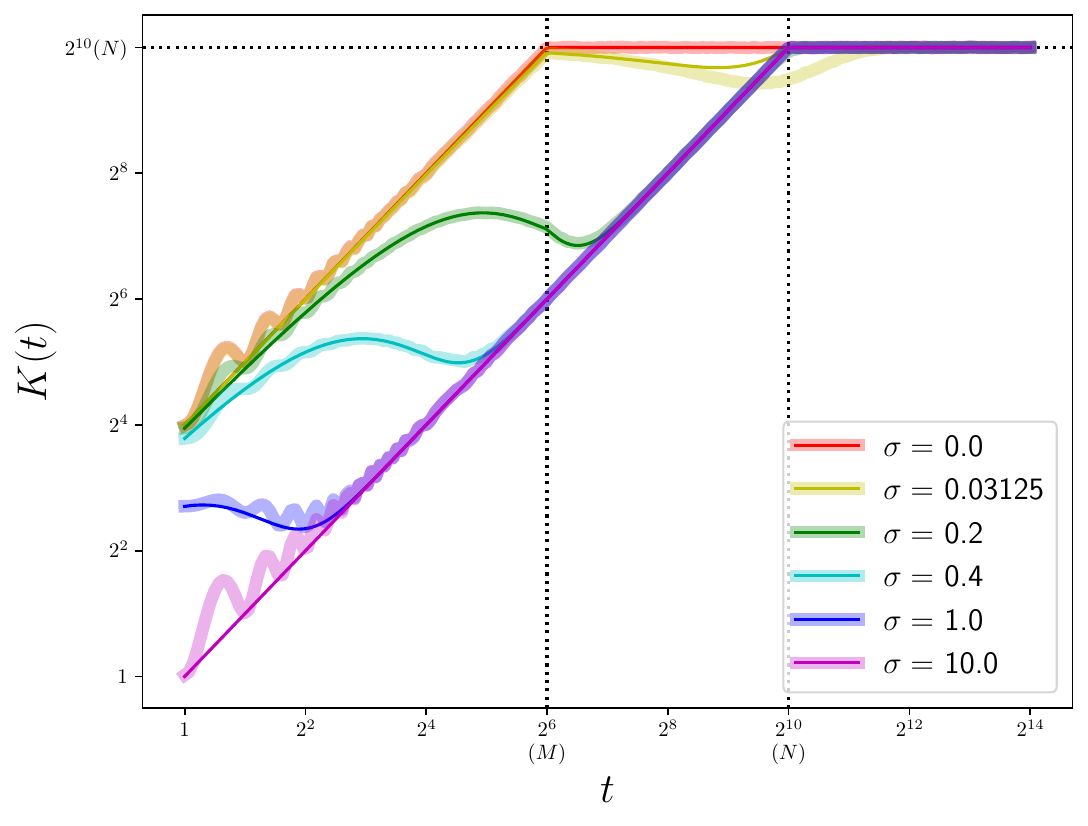}
      \caption{Comparison between the diagrammatic result for the SFF (Eq.~\ref{eq:diag_SFF}) numerics at different values of $\sigma$ for a system of 10 qubits ($q=2$, $D=6$). The thin dark curves are the theoretical results while the wider light curves are the numerical results averaged over $10^5$ realizations of the Floquet operator.}
      \label{fig:diag_SFF}
  \end{figure}

  We also calculate the SFF to leading order using the diagrammatic method for integration over the unitary group developed by Brouwer and Beenakker~\cite{Beenakker}. The details of this calculation are contained in the Supplementary Material~\cite{Sup}. With this method we find that, for times much less than the Heisenberg time, the SFF is 
  \begin{equation}\label{eq:diag_SFF}
      K(t)=e^{-\sigma^2 t}P\min(t,M)+\left(1-e^{-\sigma^2 t}\right)t+N^2\delta_{t,0}.
  \end{equation}
  This is in agreement with the result from the sigma model approach at times less than the block Heisenberg time (Eq.~\ref{eq:Kt}). Comparison of this diagrammatic result with numerics is shown in Fig.~\ref{fig:diag_SFF}. We see that there is good agreement at times much less than the Heisenberg time. The oscillations in the numerical results at early times are a result of the finite system size. If we naively extend the diagrammatic result to later times by saying Eq.~\ref{eq:diag_SFF} holds for $t<N$ and the SFF is at its plateau for all later times, there is still good agreement with numerics at these late times, so long as the time at which the SFF moves from the glassy result to the CUE result is not on the same order as the Heisenberg time.

 From the eigenstate correlation function  in Eq.~\ref{eq:Ct}, we can also directly study the the time evolution of the density matrix using Eq.~\ref{eq:rhot-0}:
	\begin{align}\label{eq:rhot}
		\begin{aligned}
			\left\langle \rho(t) \right\rangle
			=&
			\left(\delta_{t,0}+c_1(t) \right)\rho(0)
			+
			Nc_2(t) \rho_{\msf{th}}
			\\
			&
			+
			c_3(t)  \rho_{\msf{B}}(0)
			+
			Mc_4(t) \rho_{\msf{Bth}},
		\end{aligned}
	\end{align}
	where $c_{i}(t)$ are given by Eq.~\ref{eq:ct} to the leading order.
	Here $\rho_{\msf{th}}=\hat{1}/N$ is the thermal density matrix at infinite temperature. $\rho_{\msf{B}}(0)$ and $\rho_{\msf{Bth}}$ are defined separately as
	\begin{align}
		\begin{aligned}
			(\rho_{\msf{B}}(0))_{nm}=&\rho_{nm}(0) \delta_{B(n),B(m)},
			\\
			(\rho_{\msf{Bth}}(0))_{nm}=&\frac{\delta_{nm}}{M}  \tr_{B(n)} \rho(0) ,
		\end{aligned}
	\end{align}
	where $\tr_{i}$ denotes the trace restricted to the $i$-th diagonal block, defined as
	$\tr_{i} A \equiv \sum_{j=1}^{M} A_{i(M-1)+j,i(M-1)+j}$.
	In particular, $\rho_{\msf{B}}(0)$ is the initial density matrix $\rho(0)$ with all off-diagonal blocks set to zero, making it block diagonal with the same diagonal blocks as $\rho(0)$.
	$\rho_{\msf{Bth}}$ is the block thermal density matrix to which the system, constituted of independent chaotic blocks of equal dimensions, will eventually relax.
	
	In the limits $\sigma \rightarrow \infty$ and $\sigma \rightarrow 0$, the time evolution of the current model's density matrix $\left\langle \rho(t) \right\rangle$ agrees, to the leading order, with that of CUE ($\rho_{\msf{CUE}}$) and block CUE ($\rho_{\msf{BCUE}}$), respectively~\footnote{More precisely, the coefficients in front of all $\rho's$ agree with those of the CUE and block CUE to the leading order},
	\begin{subequations}
		\begin{align}
			&\begin{aligned}\label{eq:rho-CUE}
				\left\langle 
				\rho_{\msf{CUE}}(t)
				\right\rangle
				=&
				\frac{K_N(t)-1}{N^2-1}
				\rho(0)
				+
				\frac{N^2-K_N(t)}{N^2-1}
				\rho_{\msf{th}},
			\end{aligned}
			\\
			& \begin{aligned}\label{eq:rho-BCUE}
				\left\langle 
				\rho_{\msf{BCUE}}(t)
				\right\rangle
				=&
				\delta_{t,0}
				\rho(0)
				+
				\left( \frac{K_{M}(t)-1}{M^2-1}-\delta_{t,0} \right) 
				\rho_{\msf{B}}(0)
				\\
				&+
				\frac{M^2-K_M(t)}{M^2-1}
				\rho_{\msf{Bth}}.
			\end{aligned}
		\end{align}
	\end{subequations}
	Here $K_d(t)=\min(t,d)+d^2\delta_{t,0}$ represents the  SFF for CUE of dimension $d=N,M$.
	For generic quantum chaotic systems,  Eq.~\ref{eq:rho-CUE} is also expected to hold with $K_N(t)$ replaced by the system's actual SFF, which now contains an early time non-universal slope in additional to the universal ramp and plateau.
	It is this non-universal slope which determines the detailed relaxation dynamics of the systems~\cite{Reimann-0}. 
	For the block CUE described by Eq.~\ref{eq:rho-BCUE},  the system now thermalizes within each individual sector and its density matrix eventually relaxes to $\rho_{\msf{Bth}}$. Since its sectors are uncoupled from each other, it can not fully thermalize to $\rho_{\msf{th}}$
	no matter how long one waits.

In the intermediate $\sigma$ regime, substituting Eq.~\ref{eq:ct} into Eq.~\ref{eq:rhot} shows that the term involving $\rho(0)$  in $\left\langle \rho(t) \right\rangle$  quickly becomes negligible while the term involving $\rho_{\msf{B}}(0)$ remains small compared to the other terms.
	This may no longer be the case for small $q$, where $c_1(t)$ and $c_3(t)$ are expected to decay at a rate determined by $q$, similar to the slope in the SFF.
	Although Eq.~\ref{eq:ct}  shows that the magnitudes of $c_1(t)$ and $c_3(t)$ can grow with time, their contributions are expected to remain small compared to  the rest of the terms.
	To obtain the  behaviors of $c_1(t)$ and $c_3(t)$, as well as their contribution to $\left\langle \rho(t) \right\rangle$,  at large times or for finite $q$, a non-perturbative analysis is required.
	
	By contrast, terms involving $\rho_{\msf{th}}$ and $\rho_{\msf{Bth}}$ in $\left\langle \rho(t) \right\rangle$ are non-negligible, and also comparable with each other in the intermediate $\sigma$ regime. The coefficients associated with these two terms behave such that $Nc_2(t)$ grows to 1 over time, while $Mc_4(t)$ decays to 0, with both rates given by $\sigma$.  
	This indicates a  transition from the block thermalized density matrix $\rho_{\msf{Bth}}$ to the globally thermalized one $\rho_{\msf{th}}$, with a rate determined by $\sigma$. 
	In particular, this transition can be made to be extremely slow by tuning $\sigma$.
	In the extreme limit $\sigma \rightarrow 0$, the quantum circuit could never global thermalize.
	In Fig.~\ref{fig:f2}(b), we illustrate schematically the relaxation of the expectation value $O(t)=\left\langle\tr(\rho(t)O)\right\rangle$ for an arbitrary physical observable $O$ from the block thermalized  value $O_{\msf{Bth}}=\Tr(O \rho_{\msf{Bth}})$
	to the fully thermalized value $O_{\msf{th}}=\Tr(O \rho_{\msf{th}})$. It is described by
	\begin{align}\label{eq:Ot}
		\begin{aligned}
			O(t)
			=
			e^{-\sigma^2t} O_{\msf{Bth}}
			+
			\left(
			1
			-
			e^{-\sigma^2t}
			\right)
			O_{\msf{th}}.
		\end{aligned}
	\end{align}
	In both Fig.~\ref{fig:f2}(b) and Eq.~\ref{eq:Ot}, we focus on the relaxation from block thermalized $O_{\msf{Bth}}$ to the global thermalized $O_{\msf{th}}$ and therefore have ignored contribution from terms involving $\rho(0)$ and $\rho_{\msf{B}}(0)$  in $\left\langle \rho(t) \right\rangle$, which are important for the initial relaxation from $O(0)$ to $O_{\msf{Bth}}$.
 
To fully capture the explicit time dependence of the density matrix $\rho(t)$ over the entire thermalization process, 
	particularly during the early relaxation from $\rho(0)$ to $\rho_{\msf{Bth}}$, the higher order terms in $c_{2,4}(t)$ are needed. This is because, when substituted into Eq.~\ref{eq:rhot}, their contribution becomes comparable to that of $c_{1,3}(t)$. However, based on the current leading order result, the existence of a two-step relaxation process is apparent, consisting of an early time thermalization within individual sectors from $\rho(0)$ to $\rho_{\msf{Bth}}$ followed by a later time global thermalization from $\rho_{\msf{Bth}}$ to $\rho_{\msf{th}}$.


    Eq.~\ref{eq:Ot} for the thermalization behavior and Eqs.~\ref{eq:Kt}-\ref{eq:diag_SFF} for the SFF have similar structures in that they move from the block CUE result for $t\ll\sigma^{-2}$ to the CUE result for $t\gg\sigma^{-2}$. This tells us that the Thouless time for the circuit, the time it take for the system to escape from a sector and fully thermalize, is $t_\text{Th}\sim\sigma^{-2}$. We find that the circuit exhibits a phase structure analogous to that of the block Rosenzweig-Porter model. In order to see sharp transitions we may reparameterize such that $\sigma^2\sim N^{-\gamma}$, making the Thouless time $t_\text{Th}\sim N^\gamma$. If the Thouless time is much smaller than the smallest possible timescale, which is $1$ for this circuit with discrete time steps, the system behaves ergodically. So the system is in the ergodic phase for $\gamma<0$. The Heisenberg time is the longest meaningful timescale for a quantum system. If the Thouless time becomes much larger than the Heisenberg time, the system will never escape a sector and thermalize. This means that the system is in the localized phase for $\gamma>1$. For $0<\gamma<1$ the system will be in the intermediate glassy phase, where thermalization occurs, but only on a timescale exponentially long in the system size. 
    
	In this paper, we introduce a Floquet random quantum circuit whose thermalization
	process can be made extremely slow by tuning a parameter of a specific quantum gate.
	The statistical properties of quasienergy eigenstates in this model are analogous to those of quantum spin glass models,
	but are distinct from those of quantum chaotic systems, as well as integrable and MBL systems. 
	We investigate the time evolution of the density matrix for this quantum circuit model in the limit of  large local Hilbert space dimensions and for times much shorter than the block Heisenberg time.
	We find that the relaxation process consists of an initial thermalization within weakly coupled sectors and a subsequent global thermalization over the entire Hilbert space.
 
	The explicit time dependence is described by Eq.~\ref{eq:rhot}, which is an ensemble averaged result. 
	To show that this result is typical \cite{Reimann-0} for individual realizations of the quantum circuit in the ensemble, it is necessary to examine the statistical fluctuations of $\rho(t)$  by computing higher order correlation functions of the quasienergy eigenstates.	
	A non-perturbative calculation is required to
    to investigate the small $q$ case and to
    fully understand
	the complete behavior of $\rho(t)$, including the early time thermalization within individual sectors and the later time dynamics around or after the block Heisenberg time.
    To address these issues, the non-perturbative treatment  by Kravtsov and Mirlin~\cite{kravtsov1994}, initially invented for studying the energy level correlation function in disordered systems, can be applied.
	We leave these investigations for future work.
 
	Due to the specific structure of the Floquet operator, which can be considered as weakly coupled blocks of equal dimensions, our model's relaxation process contains two distinct steps. More realistic quantum systems may display a more complex relaxation behavior, and could be modeled by a generalized quantum circuit whose Floquet operator exhibits a higher level of hierarchy. Generalizing the glassy quantum circuit to incorporate a richer structure in relaxation dynamics is another direction for future work.

    This research was sponsored by the Schwinger Foundation, Army Research Office under Grant Number W911NF-23-1-0241, the National Science Foundation under Grant No. DMR-203715, and  the NSF QLCI grant OMA-2120757.

	

	\bibliography{bib}

\end{document}


\title{
		Slow Relaxation in a Glassy Quantum Circuit
  \\
  Supplementary Materials
	}

    \author{Richard D. Barney}
    \affiliation{Joint Quantum Institute, Department of Physics, University of Maryland, College Park, Maryland, USA}
    	\author{Yunxiang Liao}
    \affiliation{Department of Physics, KTH Royal Institute of Technology, SE-106 91 Stockholm, Sweden}
    \author{Victor Galitski}
    \affiliation{Joint Quantum Institute, Department of Physics, University of Maryland, College Park, Maryland, USA}

	\maketitle

\section{Sigma model calculation for the eigenstate correlation function}

The correlation function $C_{nn'm'm}(\w)$ defined in Eq. 3 in the main text can be obtained from the generating function \cite{arxiv}
	\begin{align}\label{eq:Zder}
	\begin{aligned}
	\mathcal{Z}[J]
	=&
        \left\langle 
	\int_0^{2\pi} 
        \frac{d\phi}{2\pi}
	\dfrac
	{\det \left( 1-\aF e^{i\phi} U+ J^+\right) 
		\left( 1-\bF e^{-i\phi} U^{\dagger} +J^-\right) 
	}
	{\det \left( 1-\aB e^{i\phi} U\right) 
		\left( 1-\bB e^{-i\phi} U^{\dagger}\right) 
	}
	\right\rangle,
	\end{aligned}
	\end{align}
 by taking derivatives with respect to the matrix elements of sources $J^{\pm}$ and setting them to zero at the end:
 	\begin{align}\label{eq:C1}
	\begin{aligned}
	    &\bar{C}_{nn'm'm} (\aF\bF)
        \equiv 
         \dfrac{\partial^2\mathcal{Z}[J]}{\partial J^+_{n'n}\partial J^-_{mm'}}
        	\bigg\lvert_{J=0}
	=
	\sum_{t=0}^{\infty}
	C_{nn'm'm}(t)
	\left( \aF\bF\right)^t,
        \\
        &C_{nn'm'm}(\omega)
			=
			\frac{1}{2\pi}
			\left( 
			\bar{C}_{nn'm'm} (\aF\bF)
			+
			\bar{C}_{n'nmm'}^{*} (\aF\bF)
			-
			\delta_{nn'}\delta_{mm'}
			\right). 
	\end{aligned}
	\end{align}
 $\alpha$ and $\beta$ are complex numbers whose product is $\alpha\beta=e^{i\w-\eta}$. The positive infinitesimal $1/M \ll\eta\ll 1$ 
 serves as a small parameter for the following perturbative calculation,
 and broaden the 2$\pi$-periodic Dirac delta function $\delta_{2\pi}(\w)$ in Eq.~3 of the main text into a Lorentzian function $\sum_{n=-\infty}^{\infty} e^{i\w n-|n|\eta}/2\pi $ of width $\eta$.
 
 The generating function $\mathcal{Z}[J]$ can be expressed as a Gaussian superintegral, where the integration variable possesses a structure within advanced/retarded space (contributing to the part associated with forward/backward time evolution $U$/$U^{\dagger}$) and within the fermion/boson space (corresponding to the numerator/denominator). This superintegral is then converted by color-flavor transformation into a sigma model representation~\cite{Altland-rev},
 	\begin{align}\label{eq:sigma}
	\begin{aligned}
	&\mathcal{Z}[J]
	=
	\det\left((1+J^+)(1+J^-)\right)
	\left\langle 
	\int D(\tilde{Z},Z)
	\exp(-S[\TZ,Z])
	\right\rangle ,
	\\
	&S[\TZ, Z]
	=
	-
	\STr \ln (1-\tilde{Z} {Z})
	+
	\STr \ln \left( 1- \alpha\beta\tilde{Z} M_J^{+} \mathbf{U} ZM_J^{-} \mathbf{U}^{\dagger} \right),
	\end{aligned}
	\end{align}
 which is exact and applicable to any time periodic system with a Floquet operator $U$.
Here $M_{J}^{\pm}$ is a block diagonal matrix in the boson-fermion space and takes the form of $M_{J}^{\pm} =	\begin{bmatrix}
	 1  & 0
	\\
	0 & (1+J^{\pm})^{-1}
	\end{bmatrix}_{\msf{BF}}$.
$Z^{ab}_{ij}$ and $\tilde{Z}^{ab}_{ij}$ are supermatrices which act in both the boson-fermion space (labeled by superscripts $a,b=\msf{B},\msf{F}$) and the Hilbert space of the entire system (labeled by subscripts $i,j=1,2,..,N$).  They are constrained by the relations: $\tilde{Z}^{\msf{FF}}=-Z^{\msf{FF}}\,^{\dagger}$, $\tilde{Z}^{\msf{BB}}=Z^{\msf{BB}}\,^{\dagger}$, and $|Z^{\msf{BB}}\,^{\dagger}Z^{\msf{BB}}|<1$.
$\STr$ is used to indicate the supertrace over the product of boson-fermion space and Hilbert space.

For the current model in the limit of large onsite Hilbert space dimension $q$, we could perform an expansion in powers of $Z$ and $\tilde{Z}$ in $S[\TZ, Z]$ and focus on the Gaussian order fluctuations. The applicability of this perturbative treatment in the current model is analogous to that for the $q$-orbital version of the Wegner model~\cite{wegner,wegner1980} in the limit of large $q$. Additionally, the $\eta= -\re\ln(\aF\bF)$ also serves as a small parameter in this perturbative calculation~\cite{Altland-rev}. From the leading order quadratic fluctuations one obtain the smoothed eigenstate correlation function with the 2$\pi$-periodic Dirac delta function replaced with a Lorenzian function of width $\eta$.

Up to quadratic order in expansion of $Z$ and $\tilde{Z}$, the action acquires the form
	\begin{align}\label{eq:Seff2}
	\begin{aligned}
	 S_{2} [ \TZ, Z]
	=&
	\sum_{i,j,i',j'} \sum_{a,b}
	s_{a}
	\tilde{Z}_{ij}^{ab}  Z_{j'i'}^{ba} 
	\left[
	\delta_{ii'}\delta_{jj'}
	-
        \alpha\beta
        \left(\delta_{ji}-\delta_{b\msf{F}}J^+_{ji}\right)
        \left(\delta_{i'j'}-\delta_{a\msf{F}}J^-_{i'j'}\right)
	\left(
	\delta_{B(i),B(j')}
		\frac{1+Ne^{-\sigma^2}}{M(N+1)}
	+
	\frac{1-e^{-\sigma^2}}{N+1} \right) 
	\right].
	\end{aligned}
	\end{align}
Here $s_{\msf{B/F}}=\pm 1$ and we have ignored terms of higher order in source $J^{\pm}$ which are not relevant to the eigenstate correlation function under study.

This action for the quadratic fluctuations can be rewritten as
\begin{equation}\label{eq:Seff2}
\begin{aligned}
	&S_{2} [X,Y,\tilde{X},\tilde{Y}]
	=	
	\sum_{a,b=\msf{B},\msf{F}}
	\sum_{i\neq j}
	s_{a}
	\tilde{Y}_{ij}^{ab} Y_{ji}^{ba}
	+
	\frac{1}{M}
	\sum_{a,b=\msf{B},\msf{F}}
	\sum_{m=1}^{P}
	\sum_{k=0}^{M-1} s_{a} \tilde{X}^{ab}_m (k) X^{ba}_m (k) 
	\\
	&-
	\sum_{a,b}
	\sum_{m,m'=1}^{P}
	s_{a}\alpha \beta
	\left(
	\delta_{m,m'}
	\frac{1+Ne^{-\sigma^2}}{M(N+1)} 
	+
	\frac{1-e^{-\sigma^2}}{N+1} 
	\right)
	\\
	&\times
	\left( 
	\tilde{X}^{ab}_m (0) 
	-
	\delta_{bF} 
	   \frac{1}{M}
        \sum_{k=0}^{M-1}
	\tilde{X}^{aF}_m (k)  \sum_{i=1}^{M} e^{-i2\pi k i/M} J^+_{(m-1)M+i,(m-1)M+i} 
	-
	\delta_{bF} 
	\sum_{i\neq j=1}^{N} \tilde{Y}_{ij}^{aF} J^+_{ji}  
	\delta_{B(i),m}
	\right)
	\\
	&
	\times 
	\left( 
	X^{ba}_{m'} (0) 
	-
	\delta_{aF} \frac{1}{M}
	\sum_{k=0}^{M-1} X^{bF}_{m'} (k)  \sum_{i'=1}^{M} e^{i2\pi k i'/M} J^-_{(m'-1)M+i',(m'-1)M+i'} 
        -
	\delta_{aF} 
	\sum_{i'\neq j'=1}^{N} Y_{j'i'}^{bF} J^-_{i'j'}  
	\delta_{B(j'),m'}
	\right).
 \end{aligned}
\end{equation}
Here $X_m(k)$ is the Fourier transform of the diagonal matrix elements of $Z$ within the block $m$, and $Y$ represents the off-diagonal matrix elements of $Z$:
	\begin{align}
	\begin{aligned}
	&X^{ab}_m (k)
	=
	\sum_{j=1}^{M}
	Z^{ab}_{(m-1)M+j,(m-1)M+j} e^{-i2\pi k j /M},
	\qquad
	&&\tilde{X}^{ab}_m(k)
	=
	\sum_{j=1}^{M}
	\tilde{Z}^{ab}_{(m-1)M+j,(m-1)M+j} e^{i2\pi k j/M},
        \\
        &Y^{ab}_{ij}=(1-\delta_{ij})Z^{ab}_{ij},
        \qquad
        &&\tilde{Y}^{ab}_{ij}=(1-\delta_{ij})\tilde{Z}^{ab}_{ij}.
	\end{aligned}
	\end{align}
We have used the notation that $B(i)\equiv \lceil i/M\rceil$ and $b(i)\equiv i-(B(i)-1)M$.

Setting the source $J=0$ in Eq.~\ref{eq:Seff2}, we obtain the bare propagators for $Y$ and $X(k)$ with nonzero momentum $k$, 
 \begin{align}\label{eq:prop-1}
\begin{aligned}
&
 \left\langle Y_{j'i'}^{b'a'}\tilde{Y}_{ij}^{ab}  \right\rangle_0
	=
	s_a s_b\left\langle \tilde{Y}_{ij}^{ab} Y_{j'i'}^{b'a'}   \right\rangle_0
	=
	s_a
	\delta_{ii'}\delta_{jj'}\delta_{aa'}\delta_{bb'},
 \\
 		&
	\left\langle X^{b'a'}_{m'} (k')\tilde{X}^{ab}_{m} (k) \right\rangle_0
	=
	s_as_b \left\langle \tilde{X}^{ab}_{m} (k) X^{b'a'}_{m'} (k') \right\rangle_0
	=
	s_a M
	\delta_{kk'}\delta_{aa'}\delta_{bb'}
	\delta_{mm'} .
	\end{aligned}
	\end{align}
 Here the angular bracket with subscript $0$ represents the functional average with respect the quadratic action $S_2$ at $J=0$.
 For the zero momentum mode $X(k=0)$, we instead have
 \begin{align}\label{eq:prop-0}
 \begin{aligned}
 	&
		\left\langle X^{b'a'}_{m'} (0)\tilde{X}^{ab}_{m} (0) \right\rangle_0
	=
	s_as_b \left\langle \tilde{X}^{ab}_{m} (0) X^{b'a'}_{m'} (0) \right\rangle_0
	=
	\begin{cases}
        \delta_{aa'}\delta_{bb'}
        s_a M
	\dfrac{
		(1-	\alpha\beta	)
		+
		\alpha\beta	
		\frac{(1-
		e^{-\sigma^2})M}{N+1}
	}
	{
		\left( 
		1
		-
		\alpha\beta	
             \frac{1+Ne^{-\sigma^2}}{1+N}
		 \right) 
		\left( 1
		-
		\alpha\beta
		\right) 
		} ,
	& m=m',
 	\\
        \delta_{aa'}\delta_{bb'}
        s_a M
	\dfrac{
		\alpha\beta	\frac{(1-e^{-\sigma^2})M}{N+1}
	}
	{	
\left( 
		1
		-
		\alpha\beta
             \frac{1+Ne^{-\sigma^2}}{1+N}
		 \right) 
		\left( 1
		-
		\alpha\beta
		\right) 
	} ,
	&m\neq m'.
	\end{cases}
 \end{aligned}
\end{align} 

 Considering only the contribution from the quadratic fluctuations, the smoothed correlation function $\bar{C}_{nn'm'm} (\aF\bF)$ can be obtained from
 	\begin{align}\label{eq:Cstan-0}
	\begin{aligned}	
        \bar{C}_{nn'm'm} (\aF\bF)
	=&
	\left\langle 
	\left( 
	\delta_{nn'}\delta_{mm'}
	-
	\frac{\partial^2S_{2}}{\partial J^+_{n'n}\partial J^-_{mm'}}
	+
	\frac{\partial S_{2}}{\partial J^+_{n'n}}
	\frac{\partial S_{2}}{\partial J^-_{mm'}}
	-
	\delta_{nn'}
	\frac{\partial S_{2}}{\partial J^-_{mm'}}
	-
	\delta_{mm'}
	\frac{\partial S_{2}}{\partial J^+_{n'n}}
	\right)  \bigg\lvert_{J=0}
	\right\rangle_0. 
	\end{aligned}
	\end{align}
 We now evaluate each term in the expression above separately using Eqs.~\ref{eq:Seff2}, ~\ref{eq:prop-1}, and~\ref{eq:prop-0}:
 \allowdisplaybreaks
 			\begin{align}
			&\begin{aligned}
			&\left\langle \frac{\partial S_{2}}{\partial        J^+_{n'n}}\bigg\lvert_{J=0}\right\rangle_0
			=
			\delta_{nn'} 
                \sum_{a}
			\sum_{i=1}^{P}
			s_{a}\aF \beta \frac{1}{M}
			\left(
			\delta_{B(n),i}
			\frac{1+Ne^{-\sigma^2}}{M(N+1)}
			+
			\frac{1-e^{-\sigma^2}}{N+1} 
			\right)
			\left\langle \tilde{X}^{a\msf{F}}_{B(n)} (0) X^{\msf{F}a}_{i} (0) \right\rangle_0 
			=
		0,
            \end{aligned}
        \\
            &\begin{aligned}
        	&\left\langle \frac{\partial S_{2}}{\partial J^-_{mm'}}\bigg\lvert_{J=0} \right\rangle_0
			=
			\delta_{mm'} 	
			\sum_{b}\sum_{i=1}^{P}
			s_{F} \alpha 	\beta
			\frac{1}{M}
			\left(
			\delta_{i,B(m)}
                	\frac{1+Ne^{-\sigma^2}}{M(N+1)}
			+
			\frac{1-e^{-\sigma^2}}{N+1}
			\right)
			\left\langle \tilde{X}^{\msf{F}b}_{i} (0) X^{b\msf{F}}_{B(m)} (0) \right\rangle_0
			=0,
	\end{aligned}
        \\
			&\begin{aligned}
			&\left\langle 	\frac{\partial^2 S_{\msf{eff}}^{(s)}}{\partial    J^+_{n'n}\partial J^-_{mm'}}\bigg\lvert_{J=0}\right\rangle_0 
			=
                -
			(1-\delta_{nn'}) (1-\delta_{mm'}) 
			s_{F}\aF  \bF 
			\left(
			\delta_{B(n),B(m')}
			\frac{1+Ne^{-\sigma^2}}{M(N+1)}
		      +
		      \frac{1-e^{-\sigma^2}}{N+1} 
			\right)
			\left\langle 
			\tilde{Y}_{nn'}^{\msf{FF}} 
			Y_{m'm}^{\msf{FF}} \right\rangle_0
                \\
			&-
			\delta_{nn'} \delta_{mm'}
			\sum_{k=0}^{M-1} 	s_{F} \aF \bF 
			\frac{1}{M^2}
			\left( 
			\delta_{B(n),B(m)}
			\frac{1+Ne^{-\sigma^2}}{M(N+1)}
			+
			\frac{1-e^{-\sigma^2}}{N+1} 
			\right)
			\left\langle \tilde{X}^{\msf{FF}}_{B(n)} (k) X^{\msf{FF}}_{B(m)} (k)\right\rangle_0  
			e^{-i2\pi k (b(n)-b(m))/M}  
   \\
				&		=
			-
			\delta_{nn'} \delta_{mm'}\delta_{B(n),B(m)}
			(\aF\bF)^2 \frac{(1+Ne^{-\sigma^2})}{M(N+1)^2}
			\dfrac{
				(1-\aF\bF)
				\frac{1}{M}
                    (1+Ne^{-\sigma^2})
				+
				(2-\aF\bF)
					(1-e^{-\sigma^2})
			}
			{
				\left( 
				1
				-
				\aF\bF
				\frac{1+Ne^{-\sigma^2}}{1+N}
				\right) 
				\left( 1
				-
				\aF\bF
				\right) 
			} 
			\\
			&
			-
			\delta_{nn'} \delta_{mm'}
                (\aF\bF)^2
			\dfrac{
				(1-e^{-\sigma^2})^2 
			}
			{	(N+1)^2
				\left( 
				1
				-
				\aF\bF
				\frac{1+Ne^{-\sigma^2}}{1+N}
				\right) 
				\left( 1
				-
				\aF\bF
				\right) 
			} 
			-
			\delta_{nm}\delta_{n'm'}
			\aF  \bF 
			\left( 
			\delta_{B(n),B(n')}
			\frac{1+Ne^{-\sigma^2}}{M(N+1)}
			+
			\frac{1-e^{-\sigma^2}}{N+1} 
			\right),
			\end{aligned}
           \\
           &\begin{aligned}
           &	\left\langle \frac{\partial S_{2}}{\partial J^+_{n'n}}
			\frac{\partial S_{2}}{\partial J^-_{mm'}}\bigg\lvert_{J=0}
			\right\rangle_0
            =
            	\delta_{nn'}\delta_{mm'}
			\sum_{i,j=1}^{P}
			\sum_{a,b}
			s_{a}s_{F}(\alpha\beta)^2 
			\frac{1}{M^2}
			\left(
			\delta_{B(n), i}
	       \frac{1+Ne^{-\sigma^2}}{M(N+1)}
            +
            \frac{1-e^{-\sigma^2}}{N+1}
			\right)
			\\
			&\times
   		   \left(
			\delta_{B(m'), j}
			\frac{1+Ne^{-\sigma^2}}{M(N+1)}
			+
			\frac{1-e^{-\sigma^2}}{N+1}
			\right)
			\left\langle 					
			\tilde{X}^{aF}_{B(n)} (0) 
			X^{Fa}_{i} (0) 
			\right\rangle 
			\left\langle 
			\tilde{X}^{Fb}_{j} (0) 
			X^{bF}_{B(m)} (0) 
			\right\rangle 
			\\
			&+
			\sum_{i,j=1}^{P}
			(\alpha_F	\beta_F )^2
			\left(
			\delta_{B(n), i}
	\frac{1+Ne^{-\sigma^2}}{M(N+1)}
        +
        \frac{1-e^{-\sigma^2}}{N+1}
        			\right)
        			\left(
        			\delta_{B(m'), j}
        	\frac{1+Ne^{-\sigma^2}}{M(N+1)}
        +
        \frac{1-e^{-\sigma^2}}{N+1}
        			\right)
        			\left\langle 
        			X^{FF}_{i} (0) 
        			\tilde{X}^{FF}_{j} (0) 
        			\right\rangle 
			\\
			&\times
			\left[ 
			\delta_{nn'}\delta_{mm'}\frac{1}{M^2}
			\sum_{k=0}^{M-1}
			\left\langle
			\tilde{X}^{FF}_{B(n)} (k) 
			X^{FF}_{B(m)} (k) 
			\right\rangle 
			e^{-i2\pi k (b(n)-b(m))/M} 
			+
			(1-\delta_{nn'})(1-\delta_{mm'})
			\left\langle 
			\tilde{Y}_{nn'}^{FF}
			Y_{m'm}^{FF} 
			\right\rangle 
			\right] 
            \\
			&=
						\delta_{nn'}\delta_{mm'}
						\delta_{B(n),B(m)}
						(\aF\bF )^3
						\dfrac{
                            (Ne^{-\sigma^2}+1)
                                                          \left[
							(1-\aF\bF)	
                                \frac{1}{M}
							(Ne^{-\sigma^2}+1)^2
							+
                                (2-\aF\bF)
							( 1-e^{-\sigma^2})
                                (Ne^{-\sigma^2}+1)
							+
                                 (N+1)
							\left( 1-e^{-\sigma^2}\right)
       	                \right]
						}
						{
						(1+N)^3M	
                            \left( 
							1
							-
							\aF\bF
							\frac{1+Ne^{-\sigma^2}}{1+N}
							\right)^2 
							\left( 1
							-
							\aF\bF
							\right) 
						}  
      \\
      	&+\delta_{nn'}\delta_{mm'}
			(\aF\bF )^3
			\dfrac{
                    \left( 1-e^{-\sigma^2}\right)^2
				\left[
					(1-	\aF\bF)	(Ne^{-\sigma^2}+1)		
							+
                    (N+1)
				\right] 
						}
					{(N+1)^3
							\left( 
							1
							-
							\aF\bF
							\frac{1+Ne^{-\sigma^2}}{1+N}
							\right)^2 
							(
							1
							-
							\aF\bF
							)^2}
			\\
						&+
						\delta_{nm}\delta_{n'm'} \delta_{B(n), B(n')}
						(\aF\bF)^2
						\dfrac{
							(Ne^{-\sigma^2}+1)^2 
						}
						{
                                M(N+1)^2
							\left( 
							1
							-
							\aF\bF
							\frac{1+Ne^{-\sigma^2}}{1+N}
							\right) 
						}
						\\
						&
						+
						\delta_{nm}\delta_{n'm'} 
						(\aF\bF)^2
						\dfrac{
                             \left( 1-e^{-\sigma^2}\right)
                                \left[
                                (1
							-
							\aF\bF
                                )
							(Ne^{-\sigma^2}+1)			
							+
                                (N+1)
                            \right]
						}
						{(N+1)^2
							\left( 
							1
							-
							\aF\bF
							\frac{1+Ne^{-\sigma^2}}{1+N}
							\right) 
							\left( 1
							-
							\aF\bF
							\right) 
						}  .
           \end{aligned}
	\end{align}

Combining everything and using Eq.~\ref{eq:Cstan-0},  we find that the contribution from the Gaussian fluctuations assumes the form
 \allowdisplaybreaks
	\begin{align}\label{eq:Cstan}
        \begin{aligned}
        &\bar{C}_{nn'm'm} (\aF\bF)
	=\,
        \delta_{nn'} \delta_{mm'}\left(1+\bar{c}_1(\aF\bF)\right)
        +	
	\delta_{nm} \delta_{n'm'}
         \bar{c}_2(\aF\bF)
         +
        \delta_{nn'}\delta_{mm'}
	\delta_{B(n),B(m)}
         \bar{c}_3(\aF\bF)
         +
         \delta_{nm} \delta_{n'm'}
	\delta_{B(n),B(n')}
         \bar{c}_4(\aF\bF),
	\\
	&
	\bar{c}_1(\aF\bF)
        =
	(1-e^{-\sigma^2} )^2   
	\dfrac{
		 	\frac{1}{(N+1)^2}(\aF\bF)^2 
	}
	{  	
		\left( 
		1
		-
		\aF\bF
		\frac{1+Ne^{-\sigma^2}}{1+N}
		\right)^2 
		\left( 1
		-
		\aF\bF
		\right)^2 
	}  ,
		\\
	&
         \bar{c}_2(\aF\bF)
        =
        (1-e^{-\sigma^2} )
	\dfrac{
			\frac{1}{N+1}\aF\bF
	}
	{        		
		\left( 
		1
		-
		\aF\bF
		\frac{1+Ne^{-\sigma^2}}{1+N}
		\right) 
		\left( 1
		-
		\aF\bF
		\right) 
	}  ,
	\\
	&
         \bar{c}_3(\aF\bF)
         =
         \frac{1+Ne^{-\sigma^2}}{1+N}
	\dfrac{
		\frac{1}{M^2}
		(1-\aF\bF)		(\aF\bF )^2
		\frac{1+Ne^{-\sigma^2}}{1+N}
		+
		\frac{2}{(N+1)M}
			(\aF\bF )^2
		( 1-e^{-\sigma^2})
	}
	{
		\left( 
		1
		-
		\aF\bF
		\frac{1+Ne^{-\sigma^2}}{1+N}
		\right)^2 
		\left( 1
		-
		\aF\bF
		\right) 
	}  ,
	\\
	&
 \bar{c}_4(\aF\bF)
 =
        \frac{1+Ne^{-\sigma^2}}{1+N}
	\dfrac{
		\frac{1}{M}	\aF\bF
	}
	{
		\left( 
		1
		-
		\aF\bF
		\frac{1+Ne^{-\sigma^2}}{1+N}
		\right) 
	}  .
\end{aligned}
\end{align}

 From now on, we will focus on the regime where $\sigma \ll \sqrt{\ln N}$. From the expression above, one can 
 deduce that outside this regime where $\sigma \gg \sqrt{\ln N}$, the eigenstate correlation function of the current model becomes approximately that of the circular unitary ensemble, and as a result the model becomes chaotic. We then simplify the expression above using $N\gg 1$ and $e^{-\sigma^2} \gg N^{-1}$:
	\begin{align}\label{eq:Cstan}
        \begin{aligned}
       & \bar{c}_1(\aF\bF)
        =
	\dfrac{
		\frac{1}{N^2} 	(\aF\bF)^2 (1-e^{-\sigma^2} )^2
	}
	{  	
		\left( 
		1
		-
		\aF\bF
		e^{-\sigma^2}
		\right)^2 
		\left( 1
		-
		\aF\bF
		\right)^2 
	}  ,
	\\
        &
	\bar{c}_2(\aF\bF)
        =
	\dfrac{
		\frac{1}{N}	\aF\bF (1-e^{-\sigma^2} )
	}
	{        		
		\left( 
		1
		-
		\aF\bF
		e^{-\sigma^2}
		\right) 
		\left( 1
		-
		\alpha_F\beta_F
		\right) 
	}  ,
        \\
        &
	\bar{c}_3(\aF\bF)
        =
        e^{-\sigma^2}
	\dfrac{
		\frac{1}{M^2}
		(1-\aF\bF)		(\aF\bF )^2
		e^{-\sigma^2}
		+
		\frac{2}{NM}
			(\aF\bF )^2
		( 1-e^{-\sigma^2})
	}
	{
		\left( 
		1
		-
		\aF\bF
		e^{-\sigma^2}
		\right)^2 
		\left( 1
		-
		\aF\bF
		\right) 
	}  ,
        \\
                &
	\bar{c}_4(\aF\bF)
        =
	\dfrac{
		\frac{1}{M}	\aF\bF e^{-\sigma^2}
	}
	{
		\left( 
		1
		-
		\aF\bF
		e^{-\sigma^2}
		\right) 
	}  .
\end{aligned}
\end{align}
Making use of this result, we obtain the smoothed correlation function $C_{nn'm'm}(\w)$ defined by Eq.~3 in the main text but with the $2\pi$ Dirac delta function replaced with the Loranzian of width $\eta=-\re \ln(\aF\bF)$:
	\begin{align}\label{eq:Cw}
		\begin{aligned}
			&C_{nn'm'm}(\w)
			=
			\frac{1}{2\pi}
			\left( 
			\bar{C}_{nn'm'm} (e^{i\w})
			+
			\bar{C}_{n'nmm'}^{*} (e^{i\w})
			-
			\delta_{nn'}\delta_{mm'}
			\right) 
                \\
			=&
                \frac{1}{2\pi}
                \left[
                \delta_{nn'} \delta_{mm'}
                \left(1+c_1(\w)\right)
                +	
        	\delta_{nm} \delta_{n'm'}
                 {c}_2(\w)
                 +
                \delta_{nn'}\delta_{mm'}
        	\delta_{B(n),B(m)}
                 {c}_3(\w)
                 +
                 \delta_{nm} \delta_{n'm'}
        	\delta_{B(n),B(n')}
                 {c}_4(\w)
                 \right],
                 	\end{aligned}
	\end{align}
 where
	\begin{align}\label{eq:cw}
		\begin{aligned}
                &
			{c}_1(\w)
                =
                -
			\frac{			(1-e^{-\sigma^2} )^2}{N^2} 
			\dfrac{
			\left(1+ e^{-2\sigma^2}\right)\cos(\w) 
			-
			2e^{-\sigma^2} 
			}
			{  	
			\left[ 
			\left( 
			e^{-\sigma^2}
			-
			\cos(\w)
			\right)^2 
			+
			\sin^2(\w)
			\right]^2 
			\left(
			1-\cos(\w)
			\right) 
			},
                \\
                &{c}_2(\w)
                =
			-
			\frac{1}{N}	
				\dfrac{
				(1-e^{-\sigma^2} )(1+e^{-\sigma^2})
			}
			{
				\left( 
				e^{-\sigma^2}
				-
				\cos(\w)
				\right)^2
				+
				\sin^2(\w) 
			}  ,
			\\
			&
			{c}_3(\w)
                =
			2
                \frac{e^{-\sigma^2}}{MN}
			\dfrac{
                    \left\lbrace
				P
				e^{-\sigma^2}
				\left[
				\left( e^{-\sigma^2}-\cos(\w)\right)^2
				-
				\sin^2(\w) 
				\right] 
				+
				( 1-e^{-\sigma^2})		
				\left[ 
				2\left( e^{-\sigma^2}-\cos(\w)\right)
				+
				 e^{-2\sigma^2}-1
						\right]
                    \right\rbrace
			}
			{
				\left[ 
				\left( 
				e^{-\sigma^2}
				-
				\cos(\w)
				\right)^2 
				+
				\sin^2(\w)
				\right]^2 
			},
			\\
                & {c}_4(\w)
                =
			-
			2
			\frac{e^{-\sigma^2}}{M}
			\dfrac{
			e^{-\sigma^2}-\cos (\w)
			}
			{
				\left( 
				e^{-\sigma^2}
				-
				\cos(\w)
				\right)^2
				+
				\sin^2(\w) 
			}  .
		\end{aligned}
	\end{align}
We note that the contribution from the fluctuations beyond quadratic order needs to be taken into account to study the correlation function for nearby quasienergy eigenstates with small energy separation $\w \lesssim \eta$.

It is straightforward to see from its definition that this correlation function of quasienergy eigenstates $C_{nn'm'm}(\w)$ is related to the correlation function of the quasienergy:
\begin{align}\label{eq:R2}
\begin{aligned}
        R_2(\w)
        \equiv
	\left\langle 
	\sum_{\nu,\mu}
	\delta\left( \w- E_\nu+E_{\mu}\right) 
	\right\rangle 
        =
        \sum_{nm} C_{nnmm}(\w)
        =
        \frac{1}{2\pi}\left[ N^2 \left(1+c_1(\w)\right)+Nc_2(\w)+NMc_3(\w)+N c_4(\w) \right].
\end{aligned}
\end{align} 
Therefore, we can deduce from Eq.~\ref{eq:Cw} the smoothed quasienergy correlation function
\begin{align}
\begin{aligned}
                    R_2(\w)
				=&
				\frac{1} {2\pi}
			\left\lbrace 
				N^2
				+
				2(P-1)
\dfrac{
e^{-\sigma^2}
\left[ 
\left( 1+e^{-2\sigma^2}\right) \cos(\w)
-
2e^{-\sigma^2}
\right] 
}
{
	\left[ 
	\left( 
	e^{-\sigma^2}
	-
	\cos(\w)
	\right)^2
	+
	\sin^2(\w) 
	\right]^2 
}
-
\dfrac{
1
}
{
 1-\cos (\w) 
}
\right\rbrace .
\end{aligned}
\end{align}

In the time representation, the eigenstate correlation function for discrete time $t\geq 0$ assumes the form
\begin{align}\label{eq:Ct}
	\begin{aligned}
			&{C}_{nn'm'm} (t)
			=
			\delta_{nn'}\delta_{mm'} \left(\delta_{t,0}+c_1(t) \right)
			+
			\delta_{nm}\delta_{n'm'}
                c_2(t)
                +
			\delta_{nn'}\delta_{mm'}
                 \delta_{B(n),B(m)}
                 c_3(t)
                 +
			\delta_{nm}\delta_{n'm'}
                 \delta_{B(n),B(n')}
                 c_4(t),
 	\end{aligned}
\end{align}
where
\begin{align}\label{eq:ct}
	\begin{aligned}
                &
                c_1(t)
                =
				\frac{1}{N^2}
		\left[
		t(1+e^{-t\sigma^2})
		-
		\frac{1+e^{-\sigma^2}  }{1-e^{-\sigma^2} }
		(
		1
		-
		e^{-t\sigma^2}
		)
		\right] ,
			\\
                &
                c_2(t)
                =
				\frac{1}{N}
		\left(
		1
		-
		e^{-t\sigma^2}
		\right),
			\\
			&
                c_3(t)
                =
						\frac{1}{N^2}
		\left[
		\frac{2P\s}{1-\s} \left(1-\st \right)
		+
		(P^2-2P)
		t
		\st
		-P^2
		(\st-\delta_{t,0})
            \right]
            \\
		&	
            c_4(t)
            =
		\frac{1}{M} \left(\st-\delta_{t,0}\right)
		.
	\end{aligned}
\end{align}
The spectral form factor is given by
\begin{align}\label{eq:K}
\begin{aligned}
        &K(t)
        =
        \sum_{nm} C_{nnmm}(t)
        =
        N^2 \left(\delta_{t,0}+c_1(t)\right)+Nc_2(t)+NMc_3(t)+N c_4(t)
        \\
        &=
        N^2\delta_{t,0}+ t+(P-1)te^{-\sigma^2t}.
\end{aligned}
\end{align} 
We emphasize that these expressions are valid only at early times and higher order fluctuations are required around and beyond the block Heisenberg time.

In the limit $\sigma \rightarrow \infty$, the smoothed correlation functions become those of the CUE of dimension $N$:
\begin{align}
\begin{aligned}
&c_1(\w)=\frac{1}{N^2}\left(1-\frac{1}{2 \sin^2(\w/2)}\right),
&&c_1(t)=\frac{1}{N^2}(t-1+\delta_{t,0}),
\\
&c_2(\w)=-\frac{1}{N},
&&c_2(t)=\frac{1}{N}(1-\delta_{t,0}),
\\
&c_3(\w)=c_4(\w)=0,
&&c_3(t)=c_4(t)=0,
\\
&R_2(\w)=		\frac{1} {2\pi}
			\left( 
				N^2
            -
            \dfrac{
            1
            }
            {
            2 \sin^2(\w/2)
            }
\right),
&&
K(t)=N^2\delta_{t,0}+t.
\end{aligned}
\end{align}
On the other hand, in the limit $\sigma \rightarrow 0$, the smoothed correlation functions become equivalent to those of a block CUE, i.e., a block diagonal matrix with statistically independent CUE matrices of dimension $M$ as its diagonal blocks,
\begin{align}
\begin{aligned}
&c_1(\w)=c_2(\w)=0,
&&c_1(t)=c_2(t)=0,
\\
&c_3(\w)=\frac{1}{M^2}\left(1-\frac{1}{2 \sin^2(\w/2)}\right),
&&c_3(t)=\frac{1}{M^2}(t-1+\delta_{t,0}),
\\
&c_4(\w)=-\frac{1}{M},
&&c_4(t)=\frac{1}{M}(1-\delta_{t,0}),
\\
&R_2(\w)=		\frac{1} {2\pi}
			\left( 
				N^2
            -
            \dfrac{
            P
            }
            {
             2 \sin^2(\w/2)
            }
\right),
&&
K(t)=N^2\delta_{t,0}+Pt.
\end{aligned}
\end{align}

\section{Diagrammatic Calculation of the Spectral Form Factor}
In this section we calculate the SFF
\begin{equation}
    K(t)=\left\langle|\tr U^t|^2\right\rangle
\end{equation}
of the Floquet operator
\begin{equation}
    U=AV\Theta V^\dagger.
\end{equation}
We use the diagrammatic method for integration over the unitary group developed by Brouwer and Beenakker~\cite{Beenakker} to integrate over $V$. In this method the $V$ operators are represented by dotted lines connecting a filled dot to an empty dot and the $V^\dagger$ operators are represented by starred dotted lines connecting an empty dot to a filled dot. The dots correspond to operator indices. The $A$, $A^\dagger$, $\Theta$, and $\Theta^*$ operators are then represented by directed solid lines connecting dots of the same type. Calculation of the SFF involves the product of two traces, $\tr U^t$ and $\tr (U^\dagger)^t$. Each trace forms a loop consisting of alternating directed solid lines and dotted lines.

The average over $V$ has the form
\begin{equation}
    \left\langle V_{i_1j_1}\dots V_{i_mj_m}V_{i_1'j_1'}^*\dots V_{i_m'j_m'}^*\right\rangle=\sum_{\Pi,\Pi'}X_{\Pi,\Pi'}\prod_{k=1}^m\delta_{i_k i_{\Pi(k)}'}\delta_{j_k j_{\Pi'(k)}'}
\end{equation}
when the sequence $i_1'\dots i_m'$ is a permutation of $i_1\dots i_m$ and $j_1'\dots j_m'$ is a permutation of $j_1\dots j_m$. Otherwise the average vanishes. Performing the average requires summing over each of the possible permutations $\Pi$ and $\Pi'$. The coefficients $X_{\Pi,\Pi'}$ turn out to only depend on the cycle structure of the permutation $\Pi^{-1}\Pi'$ \cite{Samuel1980}. This allows us to write the coefficients as $X_{c_1,\dots,c_k}$, where the $c_1,\dots,c_k$ are the lengths of the cycles in the factorization of $\Pi^{-1}\Pi'$.

Each permutation is represented diagrammatically by connecting each filled (empty) dot connected to a dotted line to a filled (empty) dot connected to a starred dotted line with a solid undirected line (colored red in our figures for convenience). This means that at time $t$ there are $[(2t)!]^2$ permutations to sum over to calculate the SFF. This would quickly make the calculation intractable, but we can make progress by considering only the leading order diagrams.

To determine the contribution of each diagram we consider the U-cycles and T-cycles that are formed. The T-cycles are the loops of alternating directed solid lines and undirected solid lines. Each T-cycle corresponds to a trace of the product of the operators in the cycle following the direction of the cycle. The U-cycles are the loops of undirected solid lines and dotted lines. The length of a U-cycle is half the number of dotted lines within it. It is the lengths of the U-cycles which determine the prefactor $X_{c_1,\dots,c_k}$. These coefficients are given by the recursion relation~\cite{Samuel1980}
\begin{equation}
    NX_{c_1,\dots,c_k}+\sum_{p+q=c_1}X_{p,q,c_2,\dots,c_k}+\sum_{j=2}^kc_jX_{c_1+c_j,c_2,\dots,c_{j-1},c_{j+1},\dots,c_k}=\delta_{c_1 1}X_{c_2,\dots,c_k}.
\end{equation}
We will make use of the large $N$ expansion of these factors
\begin{gather}
    X_{c_1,\dots,c_k}=\prod_{j=1}^kX_{c_j}+O(N^{k-4t-2}),\\
    X_c=(-1)^{c-1}N^{1-2c}c^{-1}\binom{2c-2}{c-1}+O(N^{-1-2c}).
\end{gather}
Note that $c^{-1}\binom{2c-2}{c-1}$ are the Catalan numbers. In particular we will use that
\begin{gather}
    X_1\approx N^{-1},\\
    X_2\approx -N^{-3}.
\end{gather}

\begin{figure}
    \centering
    \includegraphics[width=0.5\linewidth]{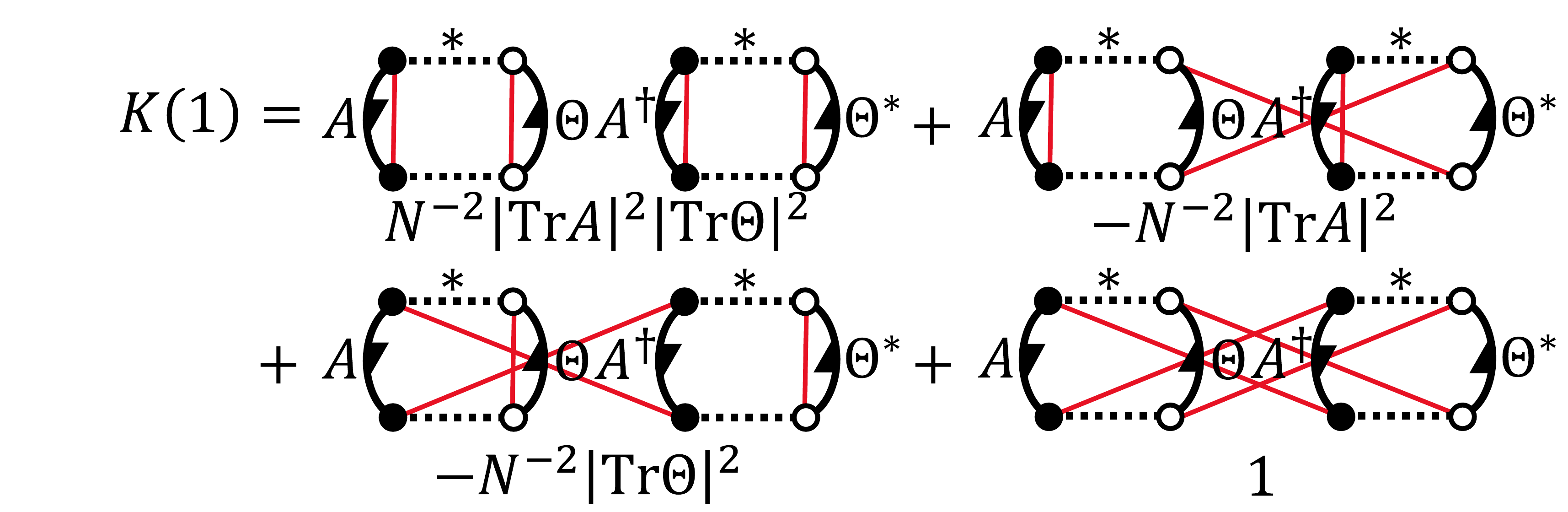}
    \caption{The diagrams contributing to the spectral form factor at $t=1$.}
    \label{fig:K_1_diags}
\end{figure}

We now examine the diagrams contributing to $K(1)$ shown in Fig.~\ref{fig:K_1_diags}. Recall that $A$ is a block diagonal matrix, with each block being a random CUE matrix with some overall phase. This means that a trace involving only $A$ and $A^\dagger$ operators will be much less than $O(N)$ if it does not come to a trace over the identity. On the other hand, traces involving only $\Theta$ and $\Theta^*$ are $O(N)$ for finite $\sigma$. This means that the second diagram in Fig.~\ref{fig:K_1_diags} is subleading and can be neglected in the large $N$ limit.

\begin{figure}
    \centering
    \includegraphics[width=0.5\linewidth]{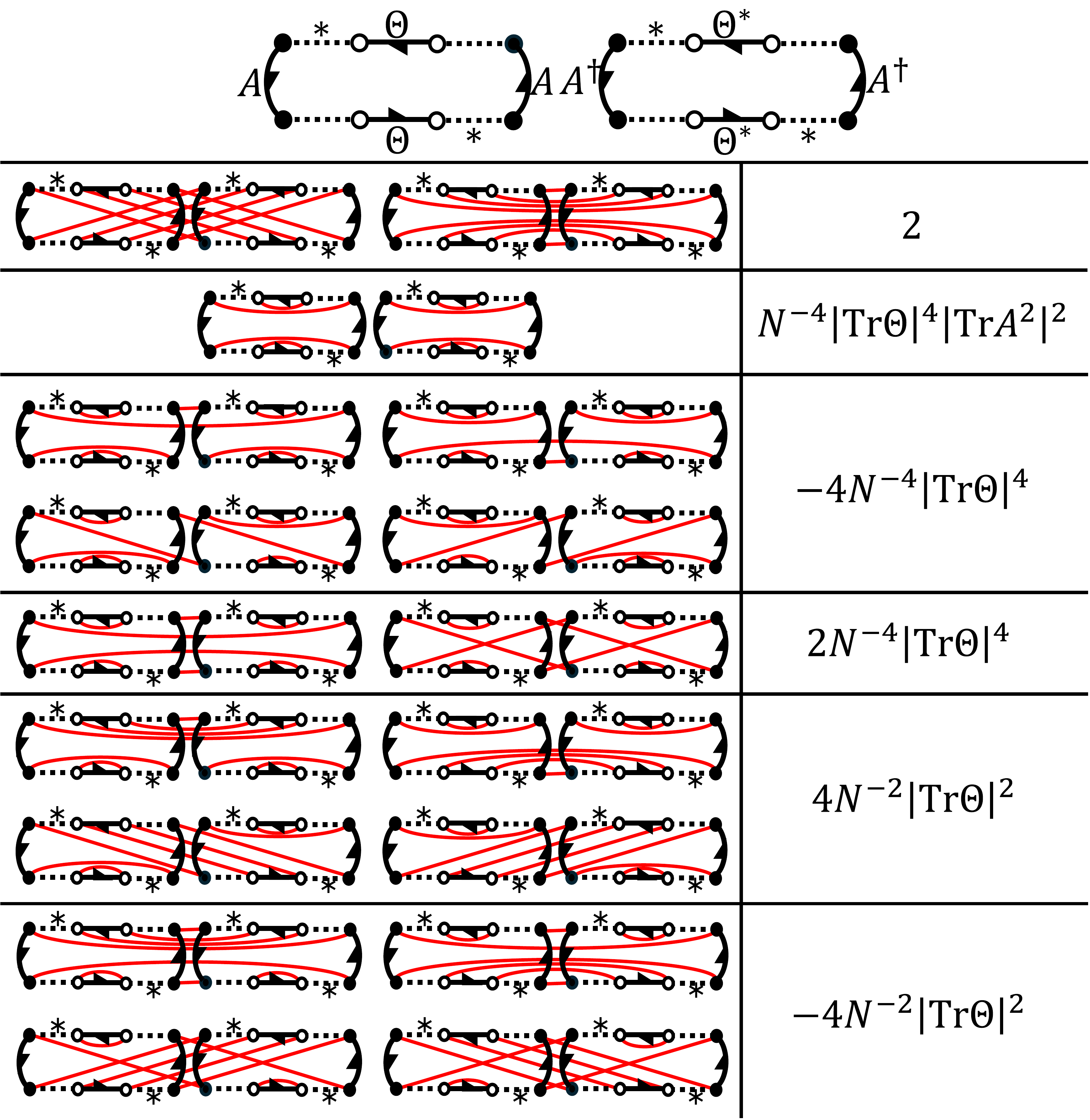}
    \caption{The leading order diagrams contributing to the spectral form factor at $t=2$. The base diagram without any contractions is shown at the top with the operators labeled. The labels are dropped in the other diagrams for clarity.}
    \label{fig:K_2_diags}
\end{figure}

Fig.~\ref{fig:K_2_diags} shows only the leading order diagrams contributing to the SFF at $t=2$. At this point we can see some patterns emerge that also hold for later times.
\begin{itemize}
    \item No T-cycles are formed that combine $A$ and $A^\dagger$ operators with $\Theta$ and $\Theta^*$ operators.
    
    \item
    The leading order diagrams do not contain U-cycles with length greater than 2.
    
    \item
    Diagrams that form T-cycles with multiple $\Theta$ or multiple $\Theta^*$ are not of leading order and can therefore be neglected at large $N$.
    
    \item
    There are always $t$ diagrams with only U-cycles of length 1 that are fully connected in the sense that each dot is connected to a dot in the other trace, as seen in the first row of Fig.~\ref{fig:K_2_diags}. The contribution from these diagrams is $t$.
    
    \item
    There is always one fully disconnected diagram of leading order, as seen in the second row of Fig.~\ref{fig:K_2_diags}. The contribution of this diagram is $N^{-2t}|\tr\Theta|^{2t}|\tr A^t|^2$.
    
    \item
    Excepting the fully connected diagrams mentioned above, each leading order diagram that forms a T-cycle consisting of a $\Theta$ and a $\Theta^*$ operator has a diagram that approximately cancels it for large $N$ that is formed by combining or dividing U-cycles. We see this in the way the diagrams in the last two rows of Fig.~\ref{fig:K_2_diags} cancel each other at large $N$. This means we can consider only diagrams in which each $\Theta$ and $\Theta^*$ is in a T-cycle by itself.
    
    \item
    For the remaining leading order diagrams there are $t\binom{t}{j}$ diagrams that connect $j$ $A$ operators to $j$ $A^\dagger$ operators for $j=1,\dots,t$. These are the diagrams shown in the third and fourth rows. For even $j$ these contributions are positive, while they are negative for odd $j$. The contribution of these diagrams is $N^{-2t}|\tr\Theta|^{2t}\sum_{j=1}^t(-1)^j\binom{t}{j}=-N^{-2t}|\tr\Theta|^{2t}t$.
\end{itemize}

With these considerations we can write the SFF as
\begin{equation}
    K(t)=N^{-2t}\left\langle|\tr\Theta|^{2t}\right\rangle\left\langle|\tr A^t|\right\rangle^2+t\left(1-N^{-2t}\left\langle|\tr\Theta|^{2t}\right\rangle\right).
\end{equation}
Considering the average over $A$ we see that
\begin{equation}
    |\tr A^t|^2=\sum_{i,j=1}^Pe^{i(\phi_i-\phi_j)}|\tr W^t|^2.
\end{equation}
Since the $\phi_i$ are uniformly distributed, only the $i=j$ terms do not vanish with averaging. $\langle|\tr W^t|^2\rangle$ is simply the dimension $M$ CUE SFF, which approaches $\min(t,M)$ for large $M$ and $t>0$. So
\begin{equation}
    \left\langle|\tr A^t|\right\rangle^2=P\min(t,M).
\end{equation}
We also note that $N^{-1}\tr\Theta$ has the form of an average, so
\begin{equation}
    \lim_{N\rightarrow\infty}N^{-2t}\left\langle|\tr\Theta|^{2t}\right\rangle=e^{-\sigma^2 t}.
\end{equation}

Putting this all together yields the SFF
\begin{equation}\label{eq:SFF_result}
    K(t)=e^{-\sigma^2 t}P\min(t,M)+\left(1-e^{-\sigma^2 t}\right)t+N^2\delta_{t,0}.
\end{equation}
The SFF may be tuned between the glassy result ($\sigma\rightarrow 0$) and the CUE result ($\sigma\rightarrow\infty$). For finite $\sigma$ the SFF will match the glassy result at early times then cross over to the CUE result at $t\sim\sigma^{-2}$. This is indicative of delayed thermalization.

This calculation of the SFF takes into account only the leading order diagrams, so it need not hold at $O(N)$ times where the lower-order diagrams may collectively make a nonvanishing contribution. We could naively extend the result to later times by saying Eq.~\ref{eq:SFF_result} holds for $t<N$ and the SFF is simply at its plateau value $N$ at all later times. Comparison with the numerical results shown in Fig.~3 of the main text does show deviation from this at late times when the thermalization time is of the same order as the plateau time. However, when this is not the case, there is good agreement between this naively extended result and numerics at all times. The oscillations in the numerical results at early times are a result of the finite system size.

\bibliography{bib.bib}